\documentclass[aip,a4paper,amsmath,amssymb,reprint,floatfix,showpacs,preprintnumbers]{revtex4-1}
\usepackage[titletoc]{appendix}
\usepackage{amsmath}
\usepackage{amssymb}
\usepackage{graphicx}
\usepackage{color}
\usepackage{subfigure}
\usepackage[sort&compress]{natbib}
\usepackage{siunitx}
\usepackage{hyperref}
\usepackage{bm}
\usepackage{soul}
\usepackage[utf8]{inputenc}
\usepackage{boondox-cal}
\usepackage{soul}
\usepackage[utf8]{inputenc}
\usepackage[T1]{fontenc}
\usepackage{mathptmx}

\newcommand{\secref}[1]{Section~\ref{#1}}
\newcommand{\n}{\mathbf{n}^{(0)}}
\newcommand{\hH}{\hat{\mathbf{H}}}

\newcommand{\vQ}{\mathbf{Q}}
\newcommand{\vn}{\hat{\mathbf{n}}}
\newcommand{\vu}{\hat{\mathbf{u}}}
\newcommand{\vone}{{\mathbb{I}}}
\newcommand{\figref}[1]{Fig.~\ref{#1}}
\newcommand{\equref}[1]{Eq.~\ref{#1}}
\newcommand{\ffigref}[2]{Fig.~\ref{#1}#2}
\newcommand{\hide}[1]{}

\renewcommand{\xi}{\~{x}_i}

\newcommand{\kBT}{k_\mathrm{B}T}                                                                                            \newcommand{\Nm}{N_\mathrm{MNP}}

\newcommand{\vH}{\mathbf{H}}
\definecolor{myorange}{RGB}{255, 69, 0} 
\definecolor{myblue}{RGB}{0, 0, 128} 
\newcommand{\val}[1]{\{= #1\}}
\newcommand{\valrange}[1]{\{\in #1\}}
\newcommand{\rep}[2]{{\textcolor{myorange}{#2}}}

\begin{document}
\title{{Non-monotonic response of a sheared magnetic liquid crystal to an external field}}
\author{Nima H. Siboni}
 \email{hamidisiboni@tu-berlin.de}
\affiliation{Institut f\"ur Theoretische Physik, Technische Universit\"at Berlin,\\  Hardenbergstra\ss e 36, 10623 Berlin, Germany}

\author{Gaurav P. Shrivastav}
\affiliation{Institute f\"ur Theoretical Physics, Technische Universit\"at Wien,\\ Wiedner Hauptstr. 8-10, 1040 Vienna, Austria}

\author{Sabine H. L. Klapp}
\affiliation{Institut f\"ur Theoretische Physik, Technische Universit\"at Berlin,\\ Hardenbergstra\ss e 36, 10623 Berlin, Germany}

\date{\today}

\begin{abstract}
  Utilizing molecular dynamics simulations,  we report a non-monotonic dependence of the shear stress on the strength of an external magnetic field ($H$) in a liquid-crystalline mixture of magnetic and non-magnetic anisotropic particles. This non-monotonic behavior is in sharp contrast with the well-studied monotonic $H$-dependency of the shear stress in conventional ferrofluids, where the shear stress increases with  $H$ until it reaches a saturation value. We relate the origin of this non-monotonicity to the competing effects of particle alignment along the shear-induced direction, on the one hand, and the magnetic field direction, on the other hand. To isolate  the role of these competing effects, we consider a {two-component} mixture composed of particles with effectively identical steric interactions, where the orientations of a small fraction, i.e.\ the magnetic ones, are coupled to the external magnetic field.
 By increasing $H$ from zero, the orientations of the magnetic particles show a Fr\'{e}ederickz-like transition and eventually start deviating from the shear-induced {orientation}, leading to an increase in shear stress. Upon further increase of $H$, a demixing of the magnetic particles from the non-magnetic ones occurs which leads to a drop in shear stress, hence creating a non-monotonic response to $H$. Unlike the equilibrium demixing phenomena reported in previous studies, the demixing observed here is neither  due to size-polydispersity nor due to a wall-induced nematic transition. Based on a simplified Onsager analysis, we rather argue that it occurs solely due to packing entropy of particles with different shear{-} or magnetic-field-induced orientations. \end{abstract}
\maketitle

\section{Introduction}
As proposed in a seminal work~\cite{brochard1970theory} by Brochard and de Gennes, doping liquid crystals (LC) with magnetic nano-particles (MNP) leads to remarkable hybrid materials whose properties can be controlled by an external magnetic field. These stimuli-responsive  materials exhibit rich self-assembly in equilibrium and show {marked} effects under external fields~\cite{rault1970ferronematics,mertelj2013ferromagnetism,mertelj2014magneto,liu2016biaxial,podoliak2012magnetite,kopvcansky2008structural}. {A key factor in determining these emergent properties is the shape of the  MNPs~\cite{kopvcansky2008structural,peroukidis2015tunable,peroukidis2015spontaneous}: Commonly, the MNPs are either {spherical} (and hence different from the LCs), or they are anisotropic with a large size disparity between the MNPs and LC particles ~\cite{chen1983observation,kopvcansky2008structural,kyrylyuk2011controlling,podoliak2012magnetite}.
{In this paper,  we investigate, on a computational basis,} {mixtures where the MNPs are identical to LCs in their anisotropy and size.} {This is a timely issue due to} recent advances in experimental realizations of such anisotropic magnetic particles~\cite{buluy2011magnetic,mertelj2013ferromagnetism,martinez2016dipolar,kredentser2017magneto}, {which has} motivated {also} a number of  analytical and numerical studies~\cite{sebastian2018director,potisk2018magneto,zarubin2018ferronematic,zarubin2018effective}. 

{The aforementioned studies~\cite{sebastian2018director,potisk2018magneto,zarubin2018ferronematic,zarubin2018effective} focus on the effect of the magnetic field on the structural and optical properties. Here, we focus on a different aspect of the response to the external field, {that is}, the mechanical response of the mixture.} A key difference between  optical and mechanical responses is that, unlike {in the} optical {case where light is transmitted through the sample without changing the structure}, {a mechanical perturbation  itself can lead to structural changes. {In particular}, when the system is sheared, {this} leads to shear-induced changes of the structure~\cite{blaak2004homogeneous,ripoll2008attractive,mokshin2008shear,mandal2012heterogeneous,shrivastav2016heterogeneous}.} We show that these shear-induced effects {combined with the effects of} the magnetic field induced effects lead to an intriguing mechanical response and structural changes.

A focus of the present study is to isolate the role of the magnetic field-induced orientation of the MNPs on the structure and rheology of {such mixtures}. To achieve {this} goal, we consider a mixture where {not only the shape and size of the MNPs and LCs are identical, but also they} interact via the same interparticular potential. {Only} the directions of the MNPs are coupled to the external magnetic field, and this coupling distinguishes the MNPs from the LCs.  {The assumed} monodispersity in size and uniformity in inter-particular interaction has several consequences{:} (i) there is no specific anchoring between the LCs and MNPs  beyond anchoring among LCs and MNPs,  (ii) polydispersity-induced phenomena, which are by themselves subject of intensive research~\cite{kikuchi2007mobile,zaccarelli2015polydispersity,heckendorf2017size,ferreiro2016spinodal,verhoeff2009liquid,speranza2002simplified}, are absent here, and (iii) {unlike Refs.~\cite{alvarez2012percolation,schmidle2012phase,sreekumari2013slow,may2016colloidal,peroukidis2016orientational,shrivastav2019anomalous}}, there is no structure formation due to the direct magnetic dipole-dipole interaction between the MNPs. These properties {of our model} enable us to isolate the role of the selectively  controlled direction of particles and study its effect on the {structure and rheology} of the whole system.

{In pure ferrofluids,} the interplay of {an} external magnetic field and shear leads to  the well-known magnetoviscous effects  (see Ref.~\cite{odenbach2004recent} and references therein): By applying the magnetic field, the viscosity of the system strongly {and monotonically} increases. This monotonic increase appears even for very dilute systems, where the dipole-dipole interactions {are negligible}~\cite{rosensweig1969viscosity,hall1969viscosity,shliomis1971effective}. {However, for mixtures containing anisotropic MNPs, such effects are} not studied, to the best of our knowledge. {Our results show that {the shear stress, and thus the viscosity,} {displays an intriguing} non-monotonic dependence on the strength of the magnetic field. We analyze this shear-stress behavior by {investigating} its correlation to the structure of the mixture at different magnetic field values.}
}

{From methodological point of view,} we use extensive nonequilibrium molecular dynamics (NEMD) simulations. Although NEMD simulations, in practice, {have} restrictions {regarding} accessible time and length scales, {they} nevertheless enable us to study the {mixture} in the presence of both shear and external magnetic field, where an analytical approach is missing. Furthermore, in NEMD simulations, no assumption on the spatial distribution of the particles is imposed, {contrary to studies \cite{sebastian2018director,potisk2018magneto} which are based on continuum theories}. In these continuum models, a homogeneous  distribution of MNPs is assumed. Our results show that such an assumption is, in fact, not fulfilled. Indeed, we find clear evidence for the formation of an inhomogeneous spatial distribution {with pronounced consequences for}  the mechanical response of the system.

In resemblance to the geometry of a rheometer, we shear the mixture by relative motion of two walls{, and obtain the shear stress directly measuring the exerted forces on the walls.} {To analyze the structure of the sample,} we use {well-established} quantities characterizing the orientational order, and also introduce a new quantity to characterize the positional distribution of MNPs across the system. Using these quantities, we {argue}  that the  non-monotonic  behavior of the stress emerges as a delicate interplay between (i) an increase in stress due to {orientational deviations of the } MNPs from the shear-induced orientation, and (ii) {a decrease in stress} due to an entropic demixing, which is caused by {differences of the  orientations of the LCs and MNPs.}

\section{Simulation setup}\label{sec:simulation-setup}

\begin{figure}
\includegraphics[width=0.25\textwidth]{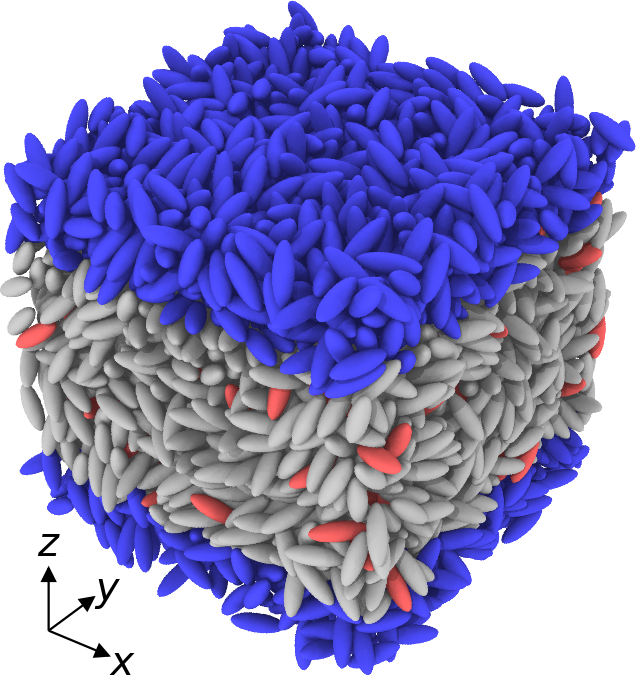}
  \caption{\label{fig:snapshot} A snapshot of the simulation setup  in absence of the shear and the external magnetic field: red, gray, and blue ellipsoidal particles represent magnetic, non-magnetic, and wall particles, respectively. The sample is in an isotropic state ($T=5$), the state {from which} the walls are created.}
\end{figure}

The simulation setup consists of ellipsoidal uniaxial Gay-Berne (GB) particles \cite{berardi1998gay}, where  a fraction of them has a permanent  magnetic point dipole embedded {in} the particle center {and} pointing along the long axis. The system is subjected to (i) shear flow, which is realized via a relative motion of two walls between which the particles are confined, and (ii) an external magnetic field.  In this section, after briefly introducing the GB parameters, the protocols for creating independent samples, shearing, and applying the external field are explained. Further, the observables and the relevant dimensionless parameters are {introduced}. A snapshot of the system, in {the} isotropic state, and in absence of the shear {flow} and magnetic field, is shown in \figref{fig:snapshot}.

For the GB potential, we adopt the  notation and parameter values used in Ref.~\cite{shrivastav2019anomalous}. Similar to Ref.~\cite{shrivastav2019anomalous} the simulation units are set such that characteristic length and energy of the GB interaction, as well as particle mass are $\sigma_0=1$ and  $\epsilon_0=1$, $m=1$, respectively. In these units, the lengths associated with semi-axes of the ellipsoids are  $\sigma_{s}=1$ and $\sigma_{l}=3$~,  for the short and the long axis of the ellipsoid, respectively. In addition to the GB interaction, the magnetic ellipsoids also interact via magnetic dipole-dipole potential. In the present work, however, we are primarily interested in the effect of the interaction of the magnetic dipoles with the external field. Therefore, the dipole-dipole interactions is set to a negligible value compared to the thermal energy. Thus, under the present conditions, thermal fluctuations {are dominant in the sense that they prevent} formation of structures due to the magnetic dipole-dipole interaction.

The protocol to create systems similar to \figref{fig:snapshot} consists of two steps: (i) independent three-dimensional configurations are created by sampling from a molecular dynamics simulation with periodic boundary conditions at high temperature, (ii) walls are created via  sudden freezing of two slabs of particles. These two steps are now explained detail.

As a first step, independent samples are created. For this purpose, an equilibrium molecular dynamics (EMD) simulation is performed with a constant number of particles, $N=4000$, at a constant volume {(a cubic box with the spatial extension $L\simeq 23$)}, at a constant temperature, $T$.  This simulation is performed with the LAMMPS~\cite{plimpton1995fast} package, where a new \texttt{fix} is introduced for embedding the point dipole moments along the longest axes of the magnetic particles.  We set the number of magnetic particles to $\Nm=200$. Newton's equation of motion for both rotational and positional degrees of freedom are integrated using the velocity-Verlet algorithm \cite{verlet1967computer,swope1982computer} with time-step $\delta t=5\times 10^{-3}$. The temperature is kept constant using a Langevin thermostat with the same parameters as in Ref. \cite{shrivastav2019anomalous}.

For the present choice of GB parameters and the number density, the equilibrium system exhibits isotropic, nematic, and smectic phases~\cite{brown1998effects}. {For the purpose of creating independent samples},  we consider the case that the equilibrium system (in the absence of shear and the magnetic field) is deep in the isotropic state. We thus set the temperature to $T=5$, which is significantly larger than the isotropic-nematic transition temperature, $T_\mathrm{IN}\simeq1.5$~\cite{brown1998effects}. To assure the independence of the configurations, the time interval between each two consecutive samplings is chosen large enough such that, on average, particles move  almost $10$ length units  in that time interval.

In a second step, we create walls for each sample for the later implementation of shear flow (see Refs.~\cite{siboni2013maintaining,varnik2003shear,hassani2018wall} for a similar strategy). To this end, the sampling at high temperature is followed by a sudden quench of the  upper and lower slabs of particles (relative to the $z$-direction), as depicted in \figref{fig:snapshot} by blue-colored ellipsoids. The MNPs frozen in the walls are replaced with the LCs, such that the walls are solely composed of nonmagnetic particles. The thickness of the walls is chosen to be larger than the GB potential cutoff in our simulation setup. These frozen slabs  serve as solid, impenetrable, and rough walls. Furthermore, as shown in Appendix~\ref{sec:app1} in \figref{fig:denistyprofile}, the roughness is large enough that the velocity profile obtained under shear does not show slip. 

 After creating walls, the system is {quenched to the desired temperature and }sheared by a relative motion of the walls along the  $x$-direction with a constant velocity, yielding a time-independent global shear rate, $\dot{\Gamma}=\Delta V_w/L$, where $\Delta V_w$ is the difference in the velocities of the two wall. One should note, the Langevin thermostat is  decoupled from particle's $x$-coordinate  to avoid the flow induced effects caused by shearing along this direction~\cite{evans1986shear}. This is a common practice  to employ  thermostating algorithms for sheared systems~\cite{thompson1989simulations,thompson1990shear,soddemann2003dissipative,zhou2005dynamic,varnik2006structural,niavarani2008slip,bolintineanu2012no}.

We now turn to the main physical quantities of interest in our study. Most important for the mechanical response is the shear stress. We calculate this quantity directly via summation of the $x$-component of the forces exerted by the liquid particles on the upper or lower wall particles, respectively,
\begin{align} \label{eq:wallforce}
\sigma_{xz}=\frac{1}{L^2}
  \sum_{i\in \mathrm{wall}}\sum_{j\in \mathrm{fluid}} \mathbf{f}_{ij,x}~.
\end{align}
In \equref{eq:wallforce}, $ \mathbf{f}_{ij,x}$ is the $x$-component of the force on particle $i$ due to particle $j$, the summation over $i$ is restricted to particles composing the upper or lower wall, and the $j$ index runs over all fluid particles. We note that by using this method of $\sigma_{xz}$ calculation, we are not relying on the assumptions behind the virial expression for pressure tensor calculation such as  homogeneity {(as discussed in ~\cite{cheung1991atomic,todd1995pressure}), or being in equilibrium~\cite{Schwabl2006}}. Knowing the shear stress, we can calculate the apparent viscosity~\cite{leslie1968some} via $\eta_a:=\sigma_{xz}/\dot{\Gamma}$~. In this study, as we keep the shear rate constant, the qualitative behaviors of the shear stress and the apparent viscosity are identical. The ensemble averages of the stress and other quantities presented here are obtained by averaging over between $15$ to $50$ independent samples.

To characterize the {orientational} structure of the system, we measure the tensorial order parameter, 
 \begin{align} \label{eq:Qtensor}
 \vQ_\alpha=\frac{1}{N_\alpha}\sum_{i=1}^{N_\alpha}\frac{1}{2} (3 \vu_{\alpha,i}\otimes\vu_{\alpha,i}-\vone)~,
\end{align}
where $\alpha\in\{\mathrm{LC,MNP}\}$ denotes the type of the particles, $N_\alpha$ is the corresponding number of particles, $\vu_{\alpha,i}$ is the unit vector along the longest axis of the $i$-th particle of type $\alpha$, and $\vone$ denotes the {second-rank unit tensor}. The nematic order parameter, $S_\alpha$, and  the direction associated with the nematic order director, $\vn_\alpha$, are obtained as the largest eigenvalue and the corresponding eigenvector of $\vQ_\alpha$. {The nematic order parameter changes in the range $[0,1]$, where $0$ corresponds to a completely random and isotropic state, and $1$ corresponds to perfect alignment of all particles. The critical value of $S$, i.e. the value at which isotopic-to-nematic transition occurs, is $S_c\simeq 0.43$}, based on Maier-Saupe mean-field theory~\cite{maier1958einfache,maier1959einfache,maier1960einfache}.

We also measure the (particle) averaged polar order parameter, $P$, {whose instantaneous value is given by}
\begin{align} \label{eq:magnetization}
P=\frac{\mu}{N_\mathrm{MNP}}\sum_{i}^{N_\mathrm{MNP}} \vu_{\mathrm{MNP},i}~.~\vn_\mathrm{MNP}~,
\end{align}
where the summation is limited only to the MNPs, and $\mu$ is the dipole moment of each MNP.

In order to obtain quantitative information on the {instantaneous} spatial distribution of MNPs and LCs, we measure the  number density profile $\rho_\alpha(z_0):=n_\alpha(z_0-\delta z/2,z_0+\delta z/2)/(L^2\delta z)$ of particle type $\alpha$, where $n_\alpha$ is the {instantaneous} number of those particles between planes $z=z_0-\delta z/2$ and $z=z_0+\delta z/2$, and $\delta z=0.25$ is the {discretization} resolution along the $z$-axis.

The studied system is characterized by the following dimensionless parameters, whose (range of) values are mentioned in the brackets:  the particle anisotropy, \mbox{$\kappa=\sigma_l/\sigma_s~\val{3}$},  the total volume fraction of \mbox{$\Phi=Nv/L^3~\val{0.34}$} where \mbox{$v=\frac{\pi}{6}\sigma_l\sigma_s^2$} represents the volume associated with a particle,  the GB energy scaled by the thermal energy, \mbox{$\beta\epsilon_0~\val{0.67}$} where \mbox{$\beta=1/(\kBT)$}, the wall-to-wall distance scaled by the particle size, \mbox{$\tilde{L}=L_c/\sigma_s~\val{14}$} where $L_c$ is the channel width,  the fraction of magnetic ellipsoids, \mbox{$x=\Nm/N~\val{0.05}$},  the energy of dipolar coupling scaled by the thermal energy, \mbox{$\lambda=\beta\mu^2/\sigma_s^3~\val{10^{-4}}$}, the dipole-external field energy scaled by the thermal energy, \mbox{$\tilde{H}=\beta\mu H~\valrange{[0,35]}$}, and the shear-induced time scaled by a  structural relaxation time, \mbox{$\dot{\Gamma}\tau_\sigma~\val{0.05}$}, where \mbox{$\dot{\Gamma}=5\times10^{-2}$} is the imposed shear rate, and \mbox{$\tau_\sigma\simeq 1$} is the time associated with decay of the stress auto-correlation in equilibrium (see \figref{fig:sacf} in Appendix~\ref{sec:sacf}).

The abovementioned choices of $\kappa$, $\Phi$, and $\beta\epsilon_0$ are such that the corresponding pure GB system exhibits isotropic, nematic, and smectic phases over a feasible temperature range~\cite{brown1998effects}. The value of $\tilde{L}$ is chosen large enough  that the wall-induced effects are negligible, based on the following two tests {which are presented in more detail in Appendix~\ref{sec:app1}}. First, we checked that for the confined system at equilibrium, $S$ is close to nematic order parameter of the same system without walls, as shown {in \figref{fig:effectofwalls} of Appendix~\ref{sec:app1}}. Second, we checked that in {the} presence of  shear,  based on density and shear rate profiles, the wall effects are not dominating the system, as depicted in \figref{fig:denistyprofile}.

Finally, the magnetic field is introduced as follows. We first let the system reach  its steady state under shear in absence of the magnetic field. We denote this point as the beginning of for measurement time, i.e.\  $t=0$. From thereon, the magnetic field is switched on and slowly increased up to a saturation value, $H_\mathrm{max}$, with a constant rate over an interval of $t_\mathrm{max}$, i.e.\  $\vH(t)=H_{\mathrm{max}}~(t/t_\mathrm{max})\hH$, where $\hH$ denotes the time-independent direction of the uniformly applied field. The {value of $H_\mathrm{max}$} is chosen such that it {allows for} complete alignment of the magnetic particles with $\hH$ {(here $\mu H_\mathrm{max}=50$). For all the simulations in this work, except {those discussed in} the Appendix \ref{sec:dHdt} where the effect $dH/dt$ is examined, $t_\mathrm{max}$ is set to $7500$ in reduced units.}

\section{Results and discussion}\label{sec:results-discussion}
In discussing our results, we first  report {our main result, that is,} the non-monotonic behavior of the stress as a function of the applied external field. Subsequently, we correlate this mechanical behavior with the structure formation of the particles in the system. {The main temperature that we consider is $T=1.5$. We deliberately chose this temperature, which is very close to the isotropic-nematic transition, to ensure high sensitivity to the applied shear and magnetic field.}

\begin{figure}[ht]
  \centering
  \includegraphics[width=0.45\textwidth]{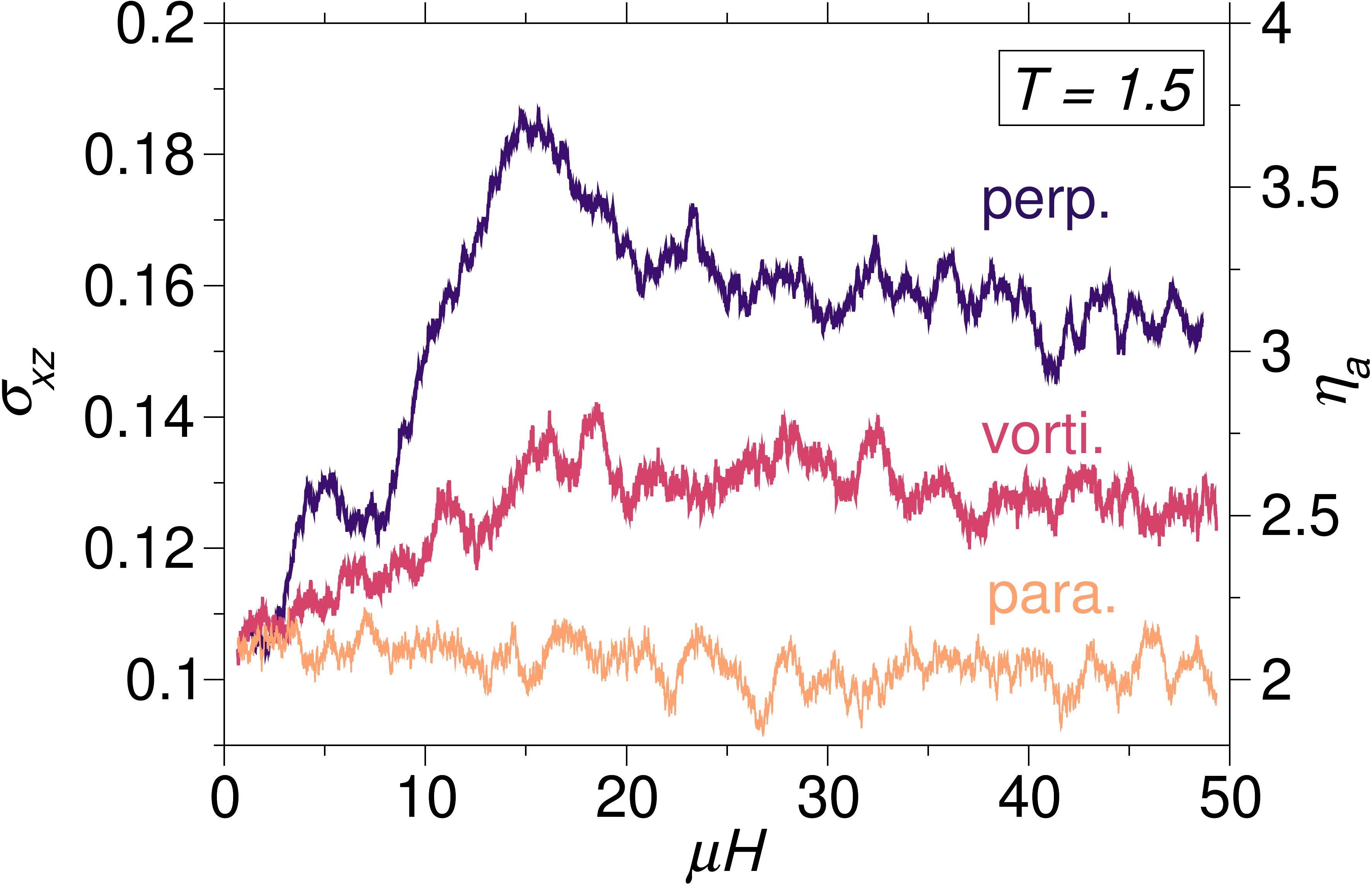}
  \caption{Shear stress as a function of the strength of an magnetic field with three different directions: (i) the  direction parallel to $\n$,  (ii) the vorticity direction, which is parallel to the $y$-axis, and (iii) the direction perpendicular to $\n$ in the $xz$-plane. The left vertical axis of the graph shows the shear stress, while the right vertical axis indicates the apparent viscosity, $\eta_a=\sigma_{xz}/\dot{\Gamma}$ with $\dot{\Gamma}=5\times 10^{-2}$~. {For all the cases, temperature is set to $T=1.5$~.}}\label{fig:different-directions} 
\end{figure}

To start with, we present in  \figref{fig:different-directions} the response of the stress to the field strength  for three different directions of $\hH$: (i) along $\n$, (ii) perpendicular to $\n$ in the $yz$-plane (vorticity direction), and (iii) perpendicular to $\n$ in the $xz$-plane {(shear plane). Here,} $\n$ indicates the nematic director of the {entire}  system in absence of the magnetic field (which is close to the direction of {the} shear {flow,} as discussed later). As shown in  \figref{fig:different-directions}, the most pronounced effect of {the} magnetic field {occurs}  in case (iii), where {we observe}, remarkably,  a non-monotonic behavior. In {the present} study, we restrict ourselves to  {case (iii)} and investigate the origins of {the observed} non-monotonicity as a function the magnetic field strength $H:=|\vH|$~.
 
\begin{figure}[ht]
 \centering 
\includegraphics[width=0.45\textwidth]{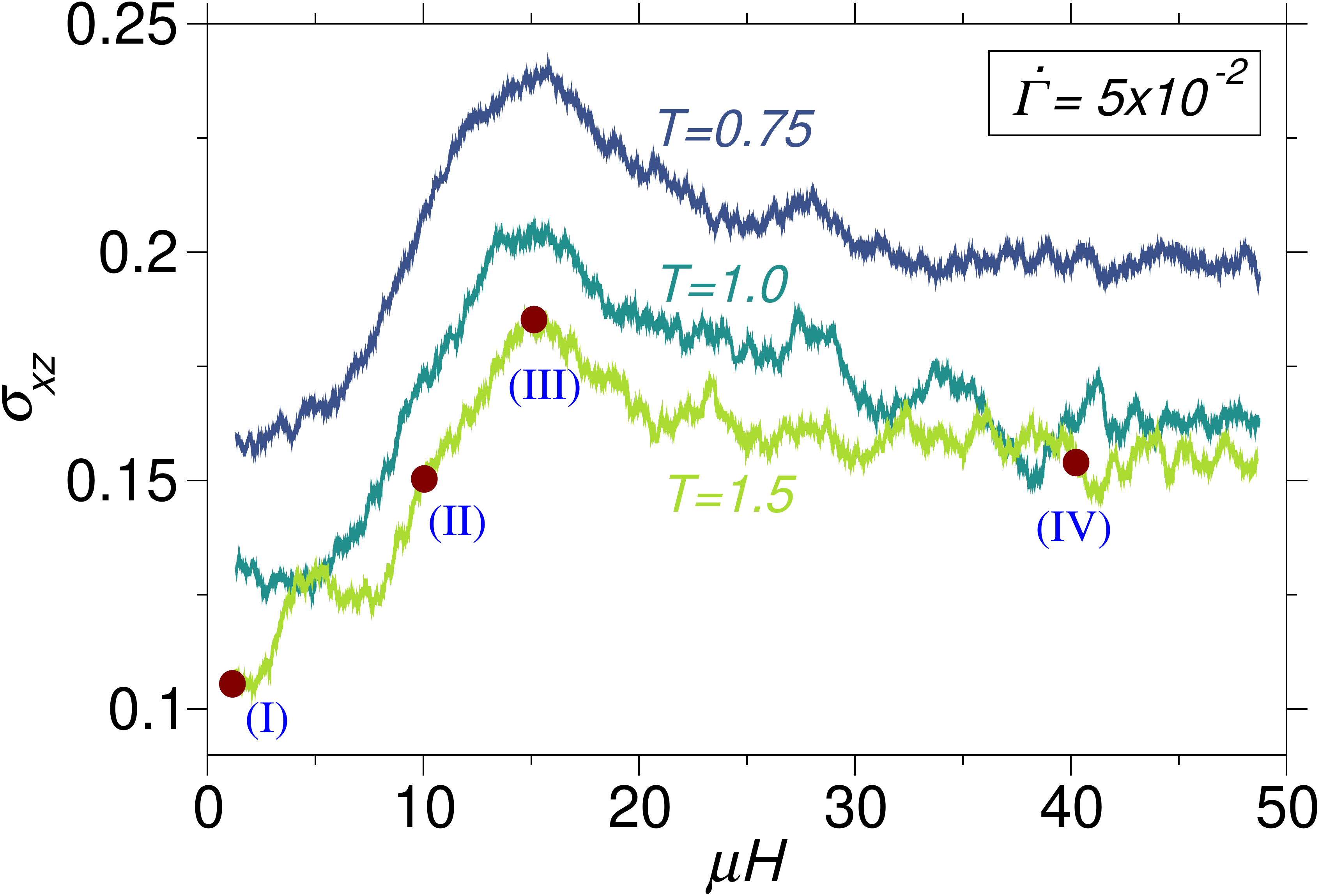}
 ~~~ \textcolor{blue}{(I)}~~~~ \includegraphics[width=0.15\textwidth]{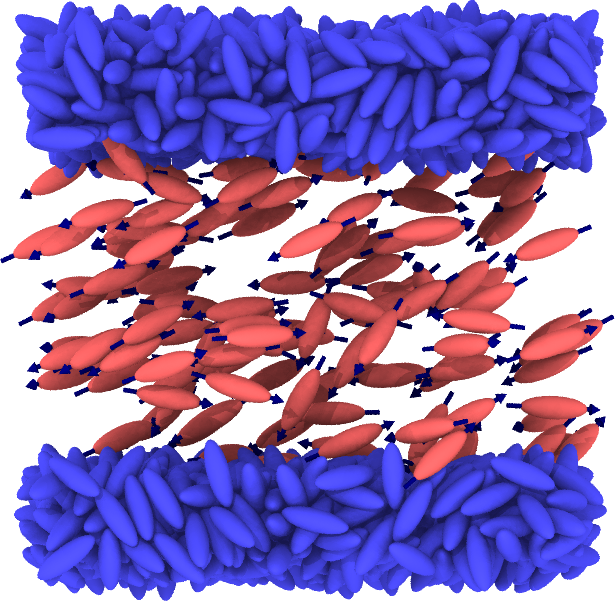}~~~
 \textcolor{blue}{(II)} \includegraphics[width=0.15\textwidth]{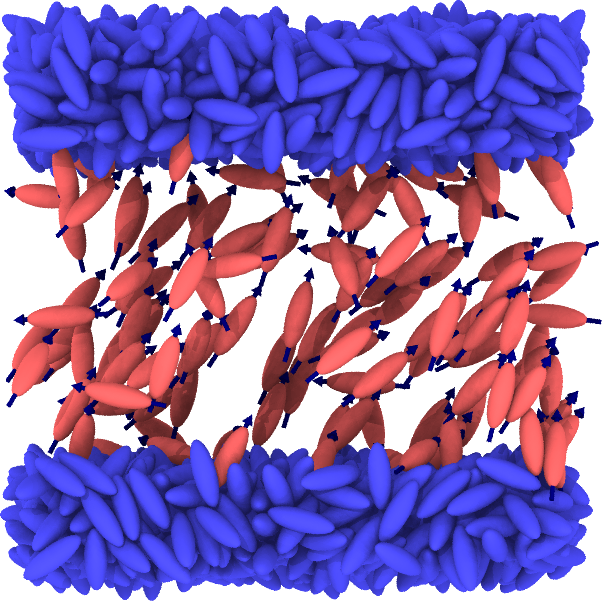}
 
 ~~~ \textcolor{blue}{(III)} \includegraphics[width=0.15\textwidth]{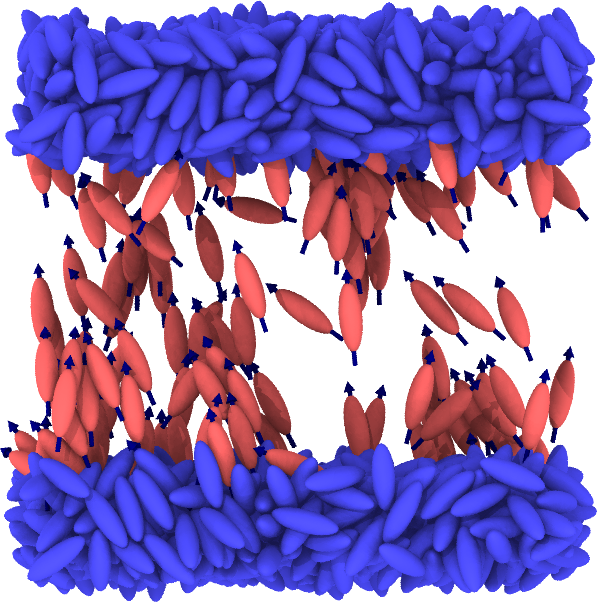}~~~
 \textcolor{blue}{(IV)}\includegraphics[width=0.15\textwidth]{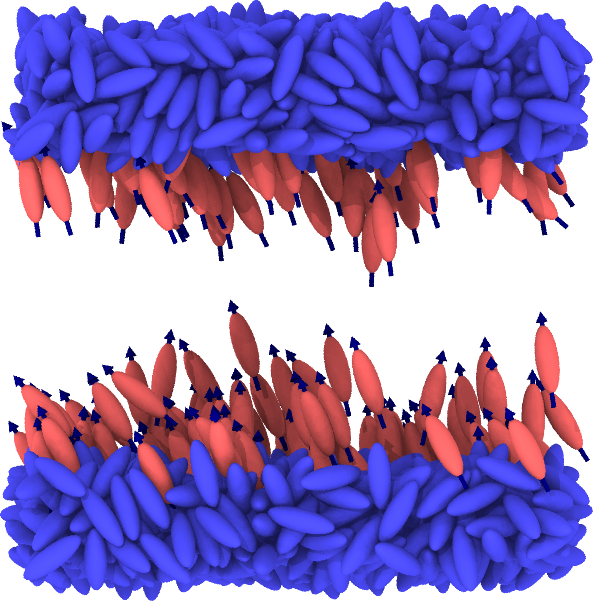}~
\caption{\label{fig:stress-vs-B} Upper panel: Shear stress as a function of the magnetic field strength at three different temperatures. Lower panel: Representative configurations associated with the four values of the field, which are indicated by bullets and the Roman numbers on the $T=1.5$ curve in the upper panel. {For all simulations, $\dot{\Gamma}$ is kept fixed at $5\times 10^{-2}$}~.}
\end{figure}

The aforementioned non-monotonic behavior is persistent, as depicted in the upper panel of \figref{fig:stress-vs-B}, at the lower temperatures $T=1.0$ and $T=0.75$, where the corresponding equilibrium system at $H=0$ is deep in the nematic state \cite{brown1998effects}. Moreover, we find the same non-monotonic behavior also for \rep{the}{} slower and faster rates of changing $H$, as shown in \figref{fig:dhdt} of Appendix~\ref{sec:dHdt}.

To shed light on the origin of this non-monotonic behavior, we now investigate the structure formation of the magnetic and non-magnetic particles at different values of the magnetic field and $T=1.5$~. The corresponding snapshots are shown in the lower panel of \figref{fig:stress-vs-B}. They reveal a significant dependence of the {orientational and translational structure} on the external field. To quantify these magnetic field induced effects, we study, {first,} the structure of the system at $H=0$, {which represents {our} reference system}. Then we show that by increasing the magnetic field for zero, the deviation of {the orientations of the} MNPs from the shear-induced direction leads to an increase of the shear stress. By further increase of $H$, the aforementioned deviation increases and eventually {causes} demixing of the MNPs and LCs. Finally, we show that this demixing is correlated with the decrease of the stress at large $H$ values.

\begin{figure}[ht]
  \centering
  \includegraphics[width=0.48\textwidth]{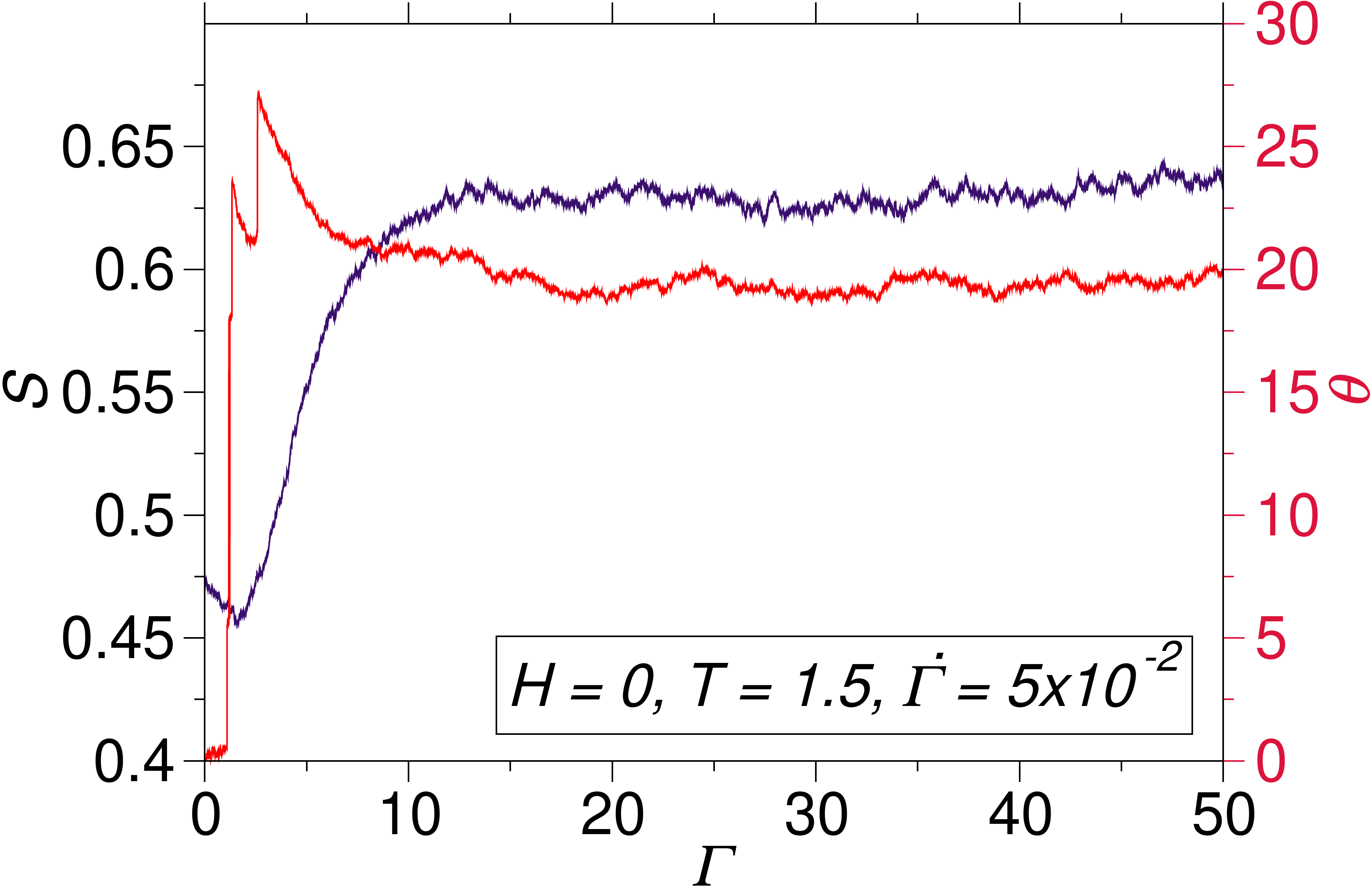}~~~~~  \caption{\label{fig:nofield}  Plots of the nematic order parameter, $S$, and the angle between the nematic order director and $x$-axis, $\theta$, {as} functions of the strain, $\Gamma$ ({in absence of the} magnetic field). Shear leads to an increase in the nematic order parameter and it also changes the orientation of the director. In absence of shear, i.e.\ at $\Gamma=0$, the director is aligned parallel to the walls and upon shearing it starts deviation from that. In steady state, we find $\theta\simeq \ang{20}$, which is close to the value reported in an earlier study~\cite{wu2007non}.  }
\end{figure}

Upon shearing the  system in absence of the magnetic field, we observe an increase in the order parameter $S_\mathrm{LC(MNP)}$ of particles, a phenomenon known as shear-induced ordering {(see Ref.~\cite{hess2004regular} and references therein)}. In \figref{fig:nofield}, the evolution of the nematic order parameter of the system  as a function applied the strain, $\Gamma$ (or, equivalently, the time interval the system is sheared), is depicted {for temperature $T=1.5$ and shear rate $\dot{\Gamma}=5\times 10^{-2}$. The results show that the system evolves from a state close to isotropic-nematic transition ($S\simeq 0.45$) to a nematic state ($S=0.65$).} We recall that at $H=0$, the LCs and the MNPs are essentially indistinguishable, and therefore we report only one $S$ for the whole system.

Applying shear not only increase the nematic order parameter, but also changes the orientation of the nematic director. In \figref{fig:nofield}, the angle between the nematic director and the shear direction (i.e. $x$-direction) is shown: the angle $\theta$ evolves from zero to its steady state value as the strain is increased. This steady state $\theta$ value, which is also known as the Leslie angle, is $\theta\simeq\ang{20}$ in our system for the previously specified parameters. The obtained value, although system specific, is close to the reported value by a previous GB study~\cite{wu2007non}.  

\begin{figure}[h!]
  \centering
 \includegraphics[width=0.43\textwidth]{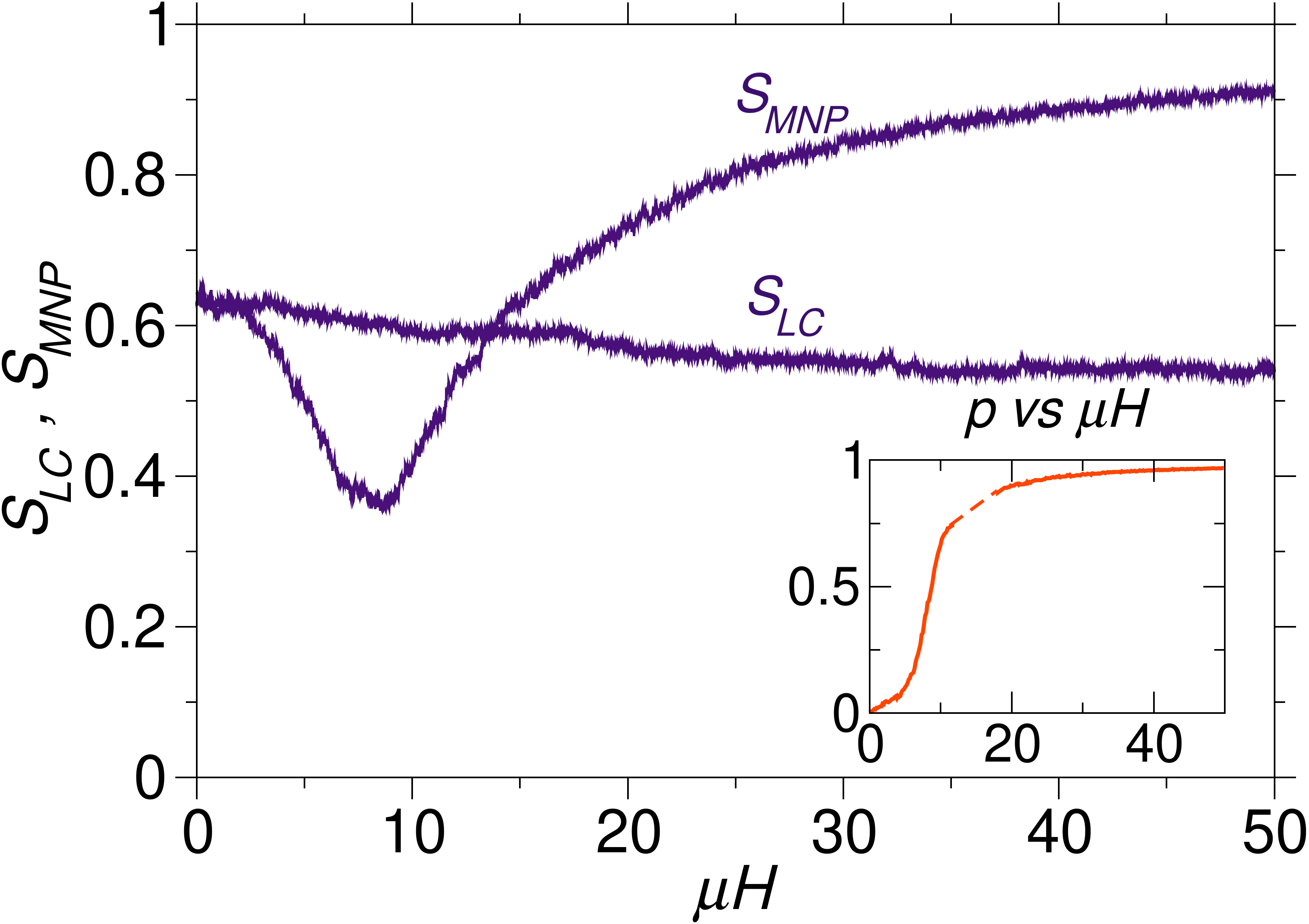}~~~~~~\\
  \vspace{3mm}
\includegraphics[width=0.45\textwidth]{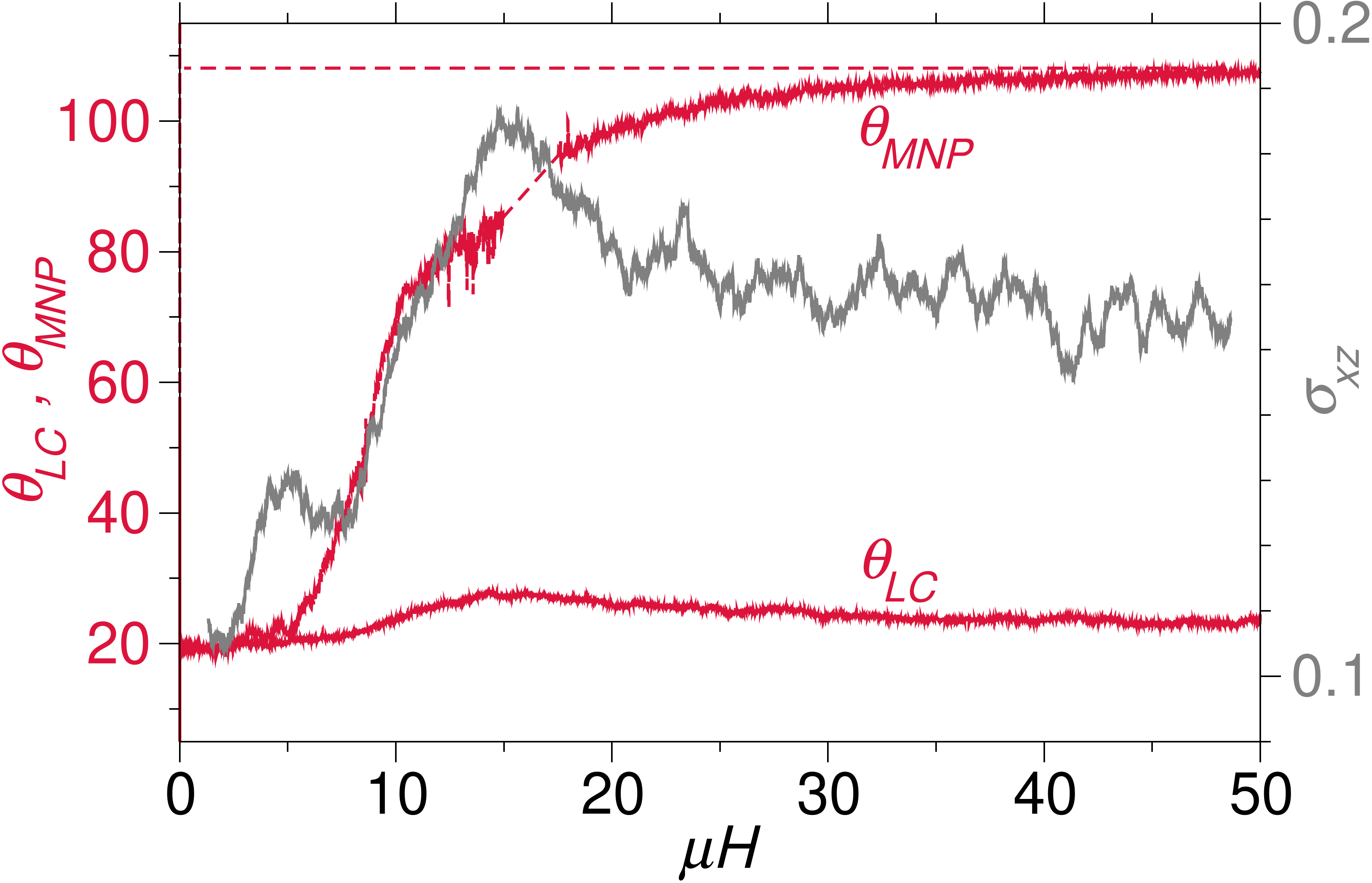}
\caption{\label{fig:nematicdirector}
  (a) Nematic order parameter for LCs and MNPs as a function of $\mu H$. The results show that, unlike  $S_\mathrm{MNP}$, $S_\mathrm{LC}$ is not affected dramatically by the external field. The inset shows that $p=\mu^{-1}P$ has a weak dependence on $\mu H$ before $\mu H \simeq 5$, in contrast to its strong dependence for larger $\mu H$ values.
  (b) The {left vertical axis} of the graph shows the angles between the nematic directors, i.e. $\vn_\mathrm{LC}$ and $\vn_\mathrm{MNP}$), with the shear direction ($x$-direction). The value of the these angles coincide at $H=0$ where the LCs and MNPs are essentially indistinguishable.  The {right vertical axis} of the graph shows the stress as a function of the external magnetic field strength. Similar to Fr\'ederickz transition, up to a {threshold value of the magnetic field} (here $\mu H_\mathrm{th}\simeq 5$) the nematic director remains undistorted, and beyond that it start deviating from the $H=0$ direction, and eventually fully aligns with the direction of the magnetic field. }
\end{figure}

Upon increasing the magnetic field strength from zero at the finite shear ($\dot{\Gamma}=5\times 10^{-2}$), the MNPs tend to align  with $\hH$, inducing a competition with the shear-induced ordering along $\n$.
As an illustration, we plot in \figref{fig:nematicdirector} the nematic order parameters of each component, as well as the angles between the director of each component, {$\vn_{\alpha}$,} and the shear direction ($x$-direction). The nematic order parameter of the MNPs, $S_\mathrm{MNP}$, decreases for small values of $H$~. This is since the MNPs now tend to align with $\hH$, which is perpendicular to the nematic director in absence of the field, i.e.\ $\n$, see \ffigref{fig:nematicdirector}{a}. Interestingly, {in the same range of $H$-values, the orientation of the nematic director is still same as for $H=0$, as seen from the behavior of $\theta_\mathrm{MNP}$} in \ffigref{fig:nematicdirector}{b}.  {Indeed,} the behavior of the angle between the MNPs director and the shear direction  is  reminiscent of a Fr\'{e}ederickz transition~\cite{freedericksz1927theoretisches,freedericksz1933forces}: Up to a threshold value of the external field , {in this case $H_\mathrm{th}\simeq 5$,} the director of magnetic particles is not deviating from $\n$. Upon further increase of the magnetic field, it suddenly starts deviating from $\n$, until it is fully aligned with the field direction, $\hat{\vH}$. In this regime, increasing the magnetic field leads to an increase of $S_\mathrm{MNP}$ along the new director, as can be seen {from} \ffigref{fig:nematicdirector}{a}. We  note that, unlike the conventional Fr\'{e}ederickz transition which is an equilibrium phenomenon, here the system is out of equilibrium. Also, in a conventional Fr\'{e}ederickz transition, the unperturbed direction is induced by the confinement, whereas here the unperturbed direction is the shear-induced direction.

The field-dependence of the ordering of magnetic particles is also reflected by the polarization,  $p=\mu^{-1}P$, {which} increases from zero by applying the magnetic field. As shown in \ffigref{fig:nematicdirector}{a}, a significant increase of $p$  occurs as the magnetic field exceeds $\mu H_\mathrm{th}$, although a slight increase is observed before this threshold.

{As is visible from \figref{fig:nematicdirector}, not only the orientational order of the MNPs, but also that of the LCs is} affected by the magnetic field, although {the} LCs {themselves} are not susceptible to the magnetic field. This is an indirect effect: The magnetic field re-orients the MNPs, which  in turn, affects the orientation of the neighboring LCs due to the  anisotropic steric interactions between the MNPs and LCs. However, considering that the MNPs form only a small fraction, the effect of the field on the LCs is small at the present condition, see \figref{fig:nematicdirector}.

In view of the competing effects of magnetic field and shear, we are {now} in a position to interpret the marked {nonmonotonic} behavior of the shear stress, $\sigma_{xz}$. Indeed, as seen from \ffigref{fig:nematicdirector}{b}, there is a clear correlation between the increase of $\sigma_{xz}$ and the misalignment of the MNPs at small to moderate values of $H$.   This can be qualitatively understood by considering that $\sigma$ is proportional to $\tau$, where $\tau$ is the relaxation time of the system. We argue that $\tau$ is increasing by applying $H$, as applying magnetic field reduces orientational freedom of the MNPs, and hence making the relaxation process less likely. Here, by the relaxation process, we refer to the atomistic mechanism behind the stress relaxation, i.e.\ going from a configuration with high stress to another configuration with a lower stress~\cite{Stillinger1995,Stillinger1982, Doliwa2003,Falk1998,Goldhirsch2002,Rabani1997,Lindemann1910,Ahn2013,siboni2015aging}. It has been argued~\cite{adam1965temperature,richert1998dynamics,mauro2009viscosity} that, the probability of such transition is proportional to  the configurational entropy of the system~\cite{adam1965temperature,richert1998dynamics,mauro2009viscosity}; roughly speaking,  it is more likely for a system to make a transition, if more states are available. In the studied system here, by confining the orientation of a fraction of particles in a particular direction via magnetic field, the configurational entropy is reduced, which leads to an increase in $\tau$ and $\sigma$.  

\begin{figure}[ht]
  \centering
\includegraphics[width=0.45\textwidth]{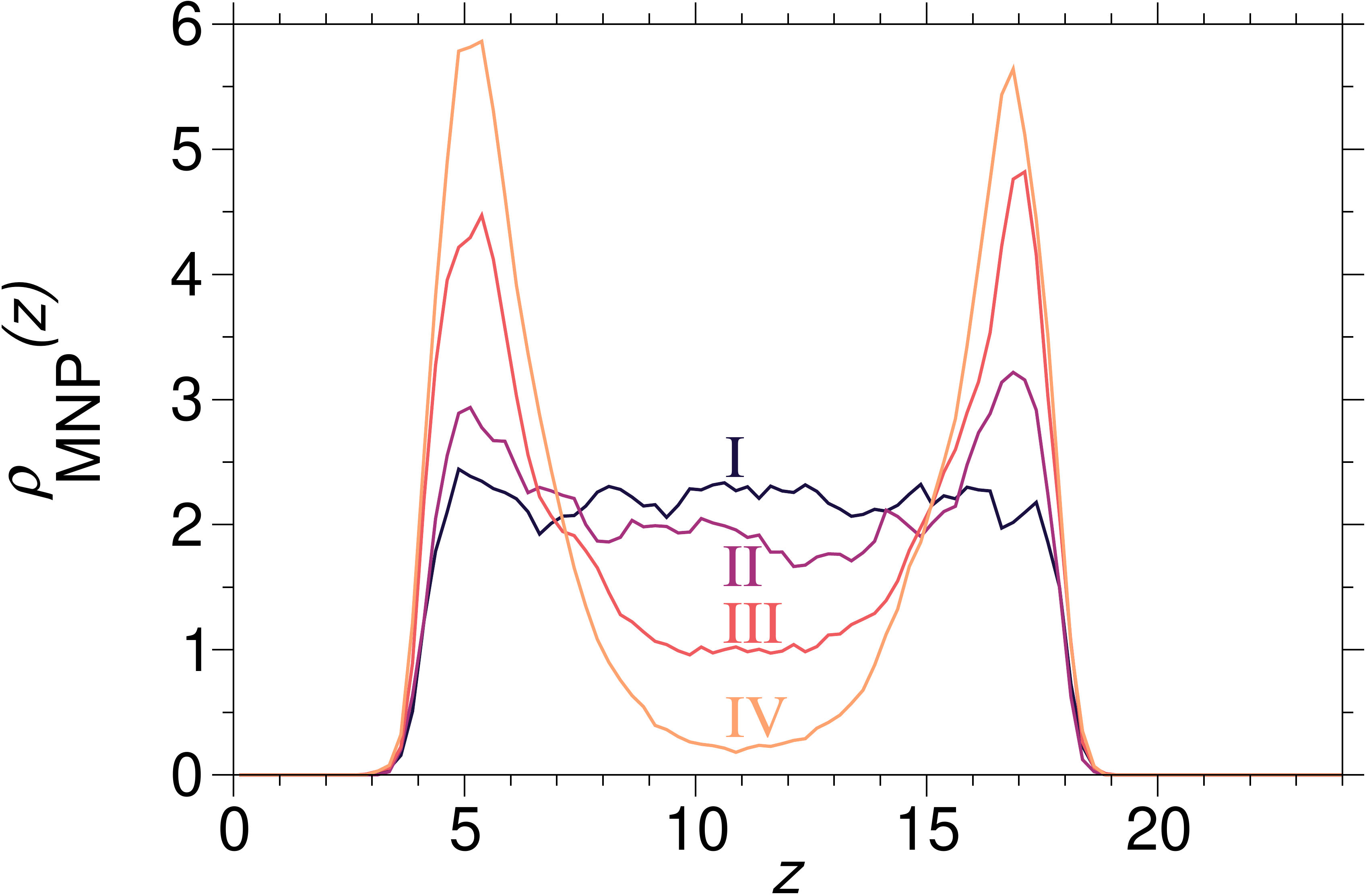}
  \caption{\label{fig:density-profiles}The density profiles of {the MNPs} at different magnetic field values. The roman numbers refer to the same numbers in \figref{fig:stress-vs-B}. {For small fields, a relatively uniform spatial distribution is obtained, in contrast to strong fields where a double-peaked profile emerges.}}
\end{figure}

So far, we have focused on the correlation between the {field-induced orientational ordering} of the magnetic particles and the shear stress. Upon further increase of the magnetic field {to the point}, where the misalignment between MNPs and LCs is close to its maximum, the shear stress shows a decrease. Interestingly, this decrease occurs simultaneously with a significant qualitative change {in} the spatial distribution of the MNPs. This qualitative change {is illustrated by}  the four representative {configurations at} different values of $\mu H$, {see} \figref{fig:stress-vs-B}: {S}tarting from a relatively homogeneous distribution of MNPs at small $H$, further increase of $H$ leads to a demixing between MNPs and LCs. {We} quantify these {structural transformations} by measuring the averaged number density profile of MNPs as a function of the external field.  The results are plotted in \figref{fig:density-profiles}{: At large field strengths (state points III and IV) one observes a pronounced double-peak structure of the MNPs density profile, reflecting the assembly of the MNPs at the walls.} This is in a qualitative contrast with the uniform spatial distribution {of the MNPs (and thus, also the LCs)} {at} small $\mu H$ values. 

{To better relate the field-induced changes of the spatial distribution and the stress behavior, we introduce} an entropy-like measure which quantifies the {degree of} inhomogeneity of the density distribution of MNPs. {Specifically, we consider the quantity}
\begin{align}\label{eq:entropylike}
  I[\tilde{\rho}]:=\int{\tilde{\rho}(z)\ln(\tilde{\rho}(z))dz}~,
\end{align}
where $\tilde{\rho}(z)$ is the MNP density, {$\rho_\mathrm{MNP}$}, with normalization  $\int_{0}^{L_c}\tilde{\rho}(z)dz=1$~.  
\begin{figure}[ht]
  \centering
\includegraphics[width=0.45\textwidth]{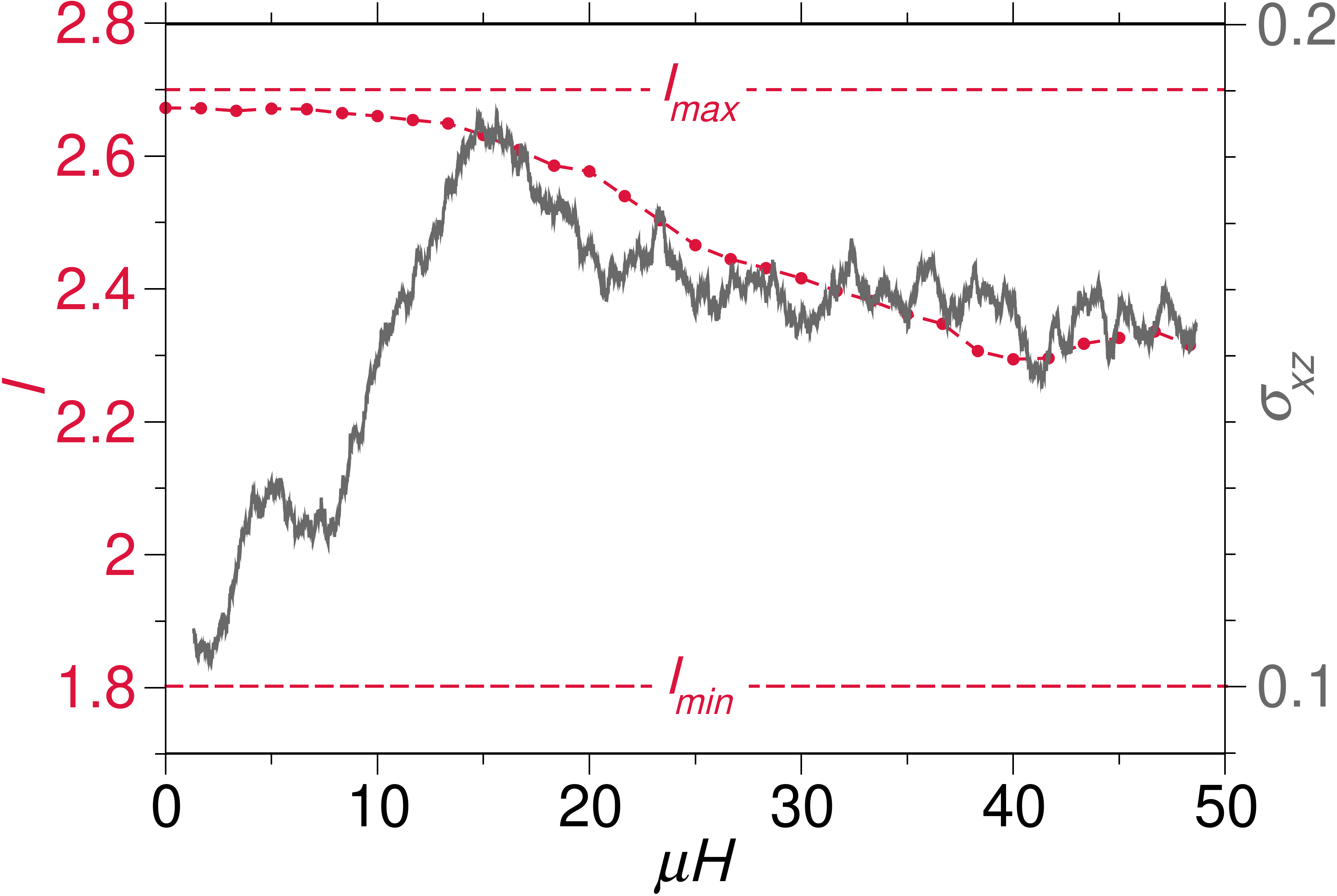}
  \caption{\label{fig:entropylike} The entropy-like quantity of \equref{eq:entropylike} as a measure of homogeneous distribution of the MNPs for different values of the external magnetic field strength. The data clearly shows that there is a transition from homogeneous spatial distribution of the MNPs to an inhomogeneous one.}
\end{figure}

The obtained $I$ as a function of the magnetic field is shown in \figref{fig:entropylike}. {For comparison, we have also indicated} the values of $I$ for two extreme cases: $I_\mathrm{max}$ {which corresponds to} an absolutely uniform distribution of MNPs, and $I_\mathrm{min}$, {which refers to} the case where all MNPs are concentrated close to walls in  a region of width $\sigma_l=3$, {the length corresponding to the large axis of the particles}. 

{For small values of $H$, $I$ remains essentially equal to its value at $H=0$, which is close to $I_\mathrm{max}$ (reflecting a nearly homogeneous distribution). Only when $H$ becomes larger than the threshold value $\mu H\simeq 15$, $I$ starts to decrease, indicating the onset of demixing.} {Moreover,} the overlay of the $H$ dependence of $I$ and $\sigma_{xy}$ {in \figref{fig:entropylike}} shows that {for larger field strengths, the stress decreases along with the $I$ decrease}.   {These results confirm the}  correlation between onset of inhomogeneity and reduction of stress.

Having established the correlation between the {demixing and the decrease} of the shear stress, there remains the question for the underlying physical mechanism. Our interpretation is as follows: The ``misalignment'' of {the} MNPs {at small fields} leads to restrictions of the {flow-induced} motion of {the} non-magnetic ones, and hence, {to an} increase of the stress. {However, this is true} only if the MNPs are dispersed {between} the LCs. As {soon as the system demixes} [{see} Figs.~(\ref{fig:density-profiles})-(\ref{fig:entropylike})], that is, at large values of $\mu H$, the MNPs are concentrated close to the walls. This {provides} a      channel for the LCs in which they can flow without (orientational {and positional}) disturbances from the MNPs.

{Clearly, a key ingredient in this line of argumentation is the field-induced orientational mismatch between the MNPs and LCs.} {Our numerical results indicate that this mismatch alone leads to a demixing-like transition:}  {We} emphasize that the interactions between all the particles are identical, so the demixing cannot be {attributed} to different shapes or interactions between the two species{, such as the demixing transitions } reported in Refs.~\cite{purdy2005nematic,varga2005nematic,bates1999nematic,ferreiro2016spinodal,helden2003direct,roth2003entropic,brader2002colloidal,Tasinkevych2006,whitmer2013,adams1998entropically,mizoshita2003fast}. We {further} {note that the demixing also occurs in bulk simulations} (where {the shear is induced} via Lees-Edward boundary condition~\cite{lees1972computer}), which emphasizes that the demixing is not due to the existence of the walls.

{We propose that the observed} demixing between  particles of different orientations can be understood as a competition between  mixing entropy and  packing entropy. {Qualitatively speaking, on the one hand, it is preferable for the system if particles with the same orientations stay close to each other as this leads to larger {(orientational)} free volumes for each particle. On the other hand, more {(positional)} configurations are available to the system if the particles are uniformly distributed in the system irrespective of their orientations. In other words, an increase in packing entropy is obtained when particles of similar orientations are neighbors, and {whereas } higher mixing entropy is reached when particles are uniformly distributed over the whole system. In our system, at small fields, where the misalignment is not yet pronounced, the mixing entropy dominates and the MNPs are uniformly distributed between the LCs. In contrast, at large fields, where the misalignments are large, the system gains entropy by bringing particles of similar orientation close to each other (and hence demixing).}

In Appendix~\ref{sec:onsager}, {by} using a simplified  Onsager analysis~\cite{onsager1949effects} for binary mixtures~\cite{lekkerkerker1984isotropic},  we argue that particle misalignments {are indeed sufficient to} to cause demixing. We show that the free energy difference between the demixed state and the fully mixed state of long, hard ellipsoids {can be written as}
\begin{align}
\label{eq:deltaFintext}
\Delta \mathcal{F}=(1-x)\ln(1-x)+x\ln(x)+2cb_\perp x(1-x)|\sin(\Theta)|~,
\end{align}
where $c$ is the number density, $b_\perp$ is the excluded volume of two ellipsoids in perpendicular configuration~\cite{onsager1949effects}, and $\Theta$ is the {angle measuring the degree of} misalignment. In the present case, $\Theta$ is essentially determined by the magnetic field. We find {from \equref{eq:deltaFintext}} that, there is a critical density {below which mixing is favored, {whereas} above it, depending on the misalignment between the directions ($\Theta$) and composition ($x$), demixing is favored.}  Although the aforementioned Onsager analysis relies on {equilibrium arguments}, {it does} suggest that  demixing {in the present nonequilibrium system} is possible.

{Experimentally, a similar} demixing has been realized {in a system of } nano-rods~\cite{van2010onsager}, where a certain degree of polydispersity is required to induce particle misalignments. In contrast to that, in our study, the demixing occurs between monodisperse particles. {This} isolates the role of orientational misalignments. {Moreover,} we want to emphasize that, in contrast to Ref.~\cite{van2010onsager}, being out-of-equilibrium is essential for the observed demixing in our system as one of the favored orientations is the shear-induced orientation.

\section{Conclusion and Outlook}   
{In this study}, we performed a NEMD study of a mixture of {anisotropic} magnetic and non-magnetic particles confined between two rough solid walls{,} {focusing on} the shear stress in the presence of an external magnetic field. {In our model, the interaction{s} between {both types of} particles are {the} same{; T}he only difference between the magnetic and non-magnetic particles is that the field acts {solely} on the direction of the magnetic particles.} Our simulation results {indicate} that the obtained shear stress depends on the strength and direction of the external field{.} More specifically,  for {a} field direction {with}in shear-plane and perpendicular to the shear-induced nematic director, we observe a non-monotonic dependence  {of shear stress and thus, the viscosity,} on the applied field strength. Such a non-monotonic behavior is in sharp contrast with the observed monotonic behavior in ferrofluids~\cite{odenbach2004recent}. {By}
 analyzing the nematic director of the LCs and MNPs, {we have found}  that the increase in the shear stress is correlated {to the} misalignment of the MNP director {relative to} the shear-induced director of the majority of particles, i.e.\ the LC particles.
 The shear stress increases up to the point that the misalignment is {sufficient to cause an entropic}  demixing between the MNPs and LCs. {The occurrence of } demixing is  also (qualitatively) predicted by a simplified Onsager analysis. Unlike previously analyzed systems where the effective entropic interaction  is due to size-polydispersity  \cite{purdy2005nematic,varga2005nematic,bates1999nematic,ferreiro2016spinodal} or shape-polydispersity \cite{helden2003direct,roth2003entropic,brader2002colloidal,Tasinkevych2006,whitmer2013,adams1998entropically,mizoshita2003fast}, in our system the {underlying mechanism is due to (competing) orientations}. {This is an interesting case of an } effective interaction where there is no difference in shape and inter-particle interaction  between the particles constituting different groups. This extends the notion of the directional entropic forces, introduced in Refs.~\cite{Damasceno2011,vanAnders2013entropically}, to a system where the particles are much simpler in their shape.

 {Given the complex response observed in the present study, one would expect even more diverse behavior when dipole-dipole interaction between the MNPs are included.} {The rheology of MNP/LC mixture in external magnetic field over a range of dipole-dipole {couplings} will be subject of further studies.}  {Indeed, {as} it is known for {systems of} pure anisotropic MNPs,} the dipole-dipole interaction can lead to self-assembled structures which can be significantly different from the chain formation observed~\cite{schmidle2012phase} in systems {of} spherical MNPs. {In particular,} for MNPs with large enough aspect ratio the neighboring particles prefer formations with anti-parallel configurations~\cite{mcgrother1998effect,alvarez2012percolation}. {We speculate that such formation of a structured phase within the liquid phase has important consequences for the rheological properties of the mixture, similar to {the} effect of {including} crystalline microstructures in amorphous bulk metallic glasses~\cite{hofmann2008designing,hays2000microstructure,guo2014shear}.} {We also expect that the external field has an important effect on the rheology: As the external field disfavors the anti-parallel dipoles, the stability of the assembled structures of MNPs is reduced~\cite{alvarez2012percolation}, which can lead to dissolving these structures in the liquid phase.}
   
 In this study, we only focused on the case where the  dipoles  are aligned with the longest axis of the ellipsoid. The anisotropic shape of the MNPs also offers the possibility of aligning the magnetic dipole along different axes of the particles. For further studies,  one can consider embedding the magnetic dipoles along the shortest axis, as done in recent experiments \cite{martinez2016dipolar}, or include an offset from the center as done for spherical particles~\cite{kantorovich2011ferrofluids,klinkigt2013cluster,morphew2016supracolloidal,steinbach2016non,yener2016self,rutkowski2017simulation}. {Depending on different embeddings,} by increasing the dipole-dipole interaction strength we speculate to find intriguing structures where an external magnetic field can play an important role in stabilizing or destabilizing them. 
\begin{acknowledgments}
{We gratefully acknowledge funding support from the Deutsche Forschungsgemainschaft (DFG) via the priority program SPP 1681.}
\end{acknowledgments}

\appendix
\section{Effects of walls}\label{sec:app1}
\begin{figure}[ht]                                                                           \includegraphics[width=0.5\textwidth]{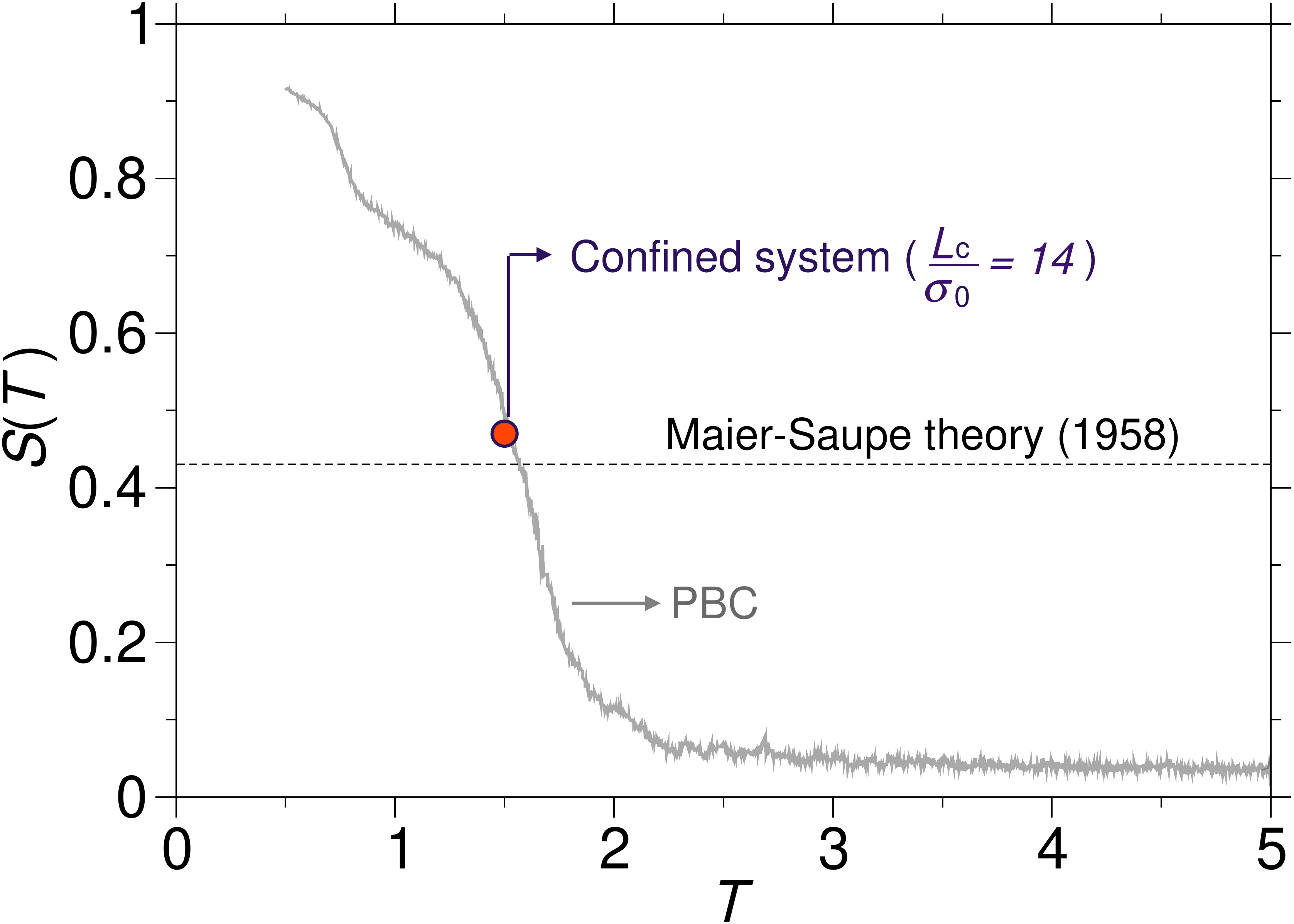}
  \caption{\label{fig:effectofwalls} Nematic order {parameter} as a function of temperature. The gray line {pertains to} the system with the full periodic boundary conditions, and the orange dot presents {the value of $S$ for the same system at $T=1.5$, after creation of the walls}, as explained in \secref{sec:simulation-setup}. The horizontal dashed line represent the approximate critical nematic order parameter \cite{maier1958einfache,maier1959einfache,maier1960einfache}, $S_c\simeq 0.43$, above which the system is considered to be in the nematic phase. {The main working temperature is chosen such that the system is very close to isotopic to nematic state transition ($S\simeq 0.45$).}}
\end{figure}

Strong confinement can lead to structural and dynamical properties which are significantly different from  bulk properties{~\cite{gruhn1997microscopic,mazza2010entropy}}. Nevertheless, if the confinement is not {severe}, {one would expect} the sample {to behave} similar to the bulk. In this appendix, we argue that although the system studied here has walls and {is} therefore confined, the wall effects are negligible at the current wall-to-wall separation.
{To this end, } we first compare the nematic order (in absence of the magnetic field and shear) as a function of temperature for two systems: the system without walls (i.e. the system with the periodic boundary conditions {in all directions}), and the same system after creating walls by freezing the wall particles, as explained in the main text {(see \secref{sec:simulation-setup})}. In \figref{fig:effectofwalls}, the nematic order of the system in presence of wall is shown over {a range of temperature{s}}. {Also indicated is} the value of $S$ at the main working temperature, i.e. $T=1.5$~, {in presence of walls{.} It is seen that the} change in $S$, as a result of introducing walls, is negligible. One should note that, as the system at $T=1.5$ is very close to its isotropic-to-nematic transition, one would expect a relatively high sensitivity of $S$ on the ambient changes, including introducing walls. Even under these conditions, introducing walls does not change $S$, which is an indication that the walls do not dominate the behavior of the system.

\begin{figure}[ht]
  \centering
\includegraphics[width=0.45\textwidth]{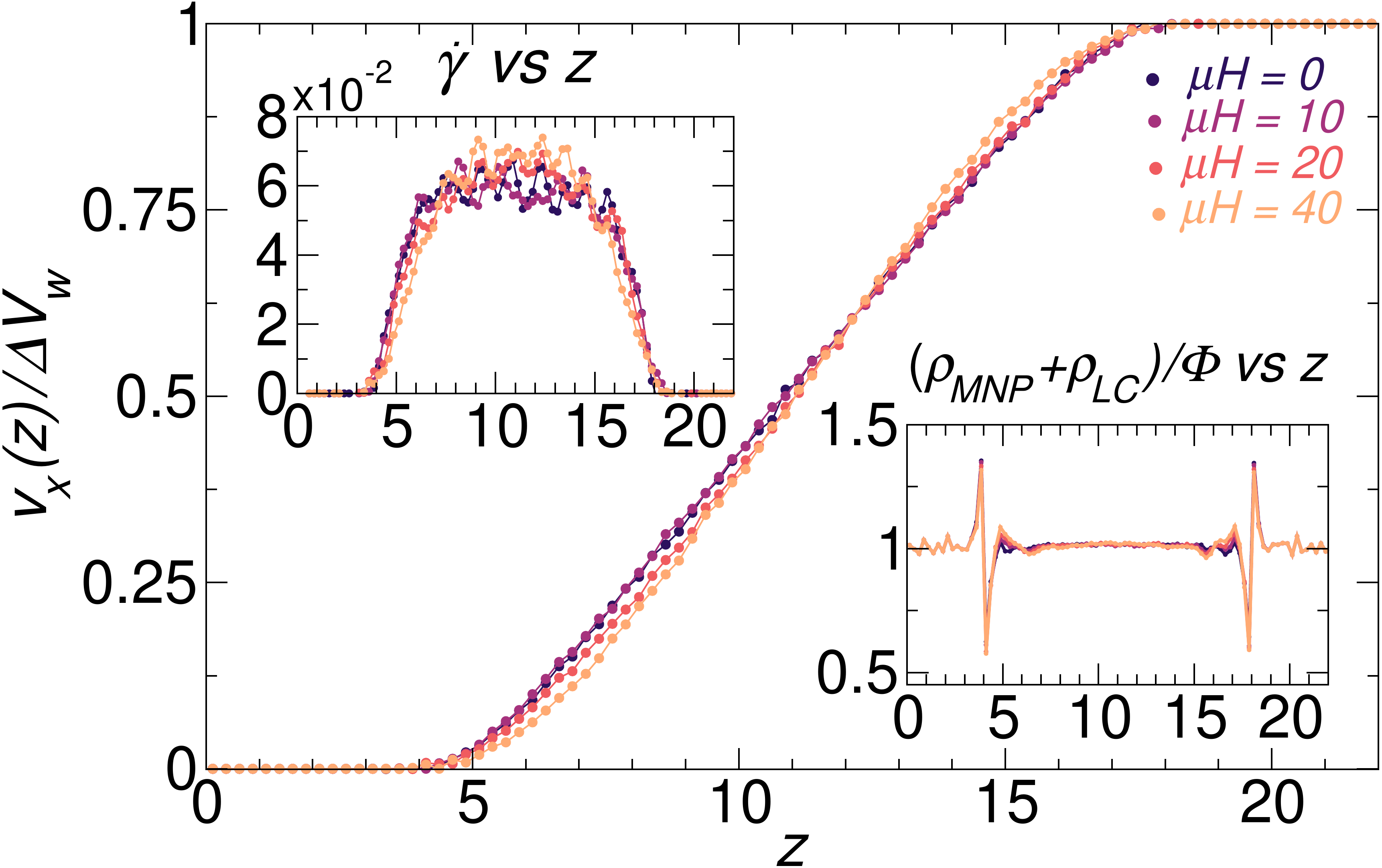}
  \caption{\label{fig:denistyprofile} The velocity profile, $v_x(z)$ which is the average velocity along $x$-direction as a function of $z$ at $T=1.5$ and four different values of $\mu H$. The local shear-rate, $\dot{\gamma}(z)=\frac{d}{dz} v_x(z)$, and the local particle density (including both MNP and LC particles) are shown in the insets. There is no indication of slip as there is no discontinuity of the shear rate across the boundary. The condition at the boundary can be approximated with no slip boundary condition with where the boundary effects are present up to $3\sigma_0$.    }
\end{figure}

{As a second point,} we check the effect of the walls in presence of shear {flow} by measuring the density and shear rate profiles across the channel (along $z$-direction) at different {strength of the external field}. The local density is obtained as described in \secref{sec:simulation-setup}, and the local shear rate is obtained by $\dot{\gamma}(z):=\frac{d}{dz} v_x(z)$, where $v_x(z_0)$ is the average  $x$-component of velocities of all particles between planes $z=z_0-\delta z/2$ and $z=z_0+\delta z/2$, and $\delta z=0.25$ is the descritization resolution along the $z$-axis. {The obtained velocity, shear rate, and density profiles are shown in \figref{fig:denistyprofile}, for $T=1.5$ and $\dot{\Gamma}=5\times 10^{-2}$~.} { The velocity profile shows that the flow is almost independent of the magnetic field strength. This is remarkable given that the composition profile and also the average orientations strongly depend on the magnetic field strength.} {The shear rate profile, {plotted} in the inset of \figref{fig:denistyprofile},} shows that there is no slip close to the walls, and {that} the wall effects on the dynamics (here the local shear rate) {reach} from the wall into the bulk of the system {over a length of about} $\simeq 3\sigma_0$. The same is  valid for the local density of all  particles.

\section{Dependence of the shear stress {on the field protocol}}
\label{sec:dHdt}
\begin{figure}
\includegraphics[width=0.45\textwidth]{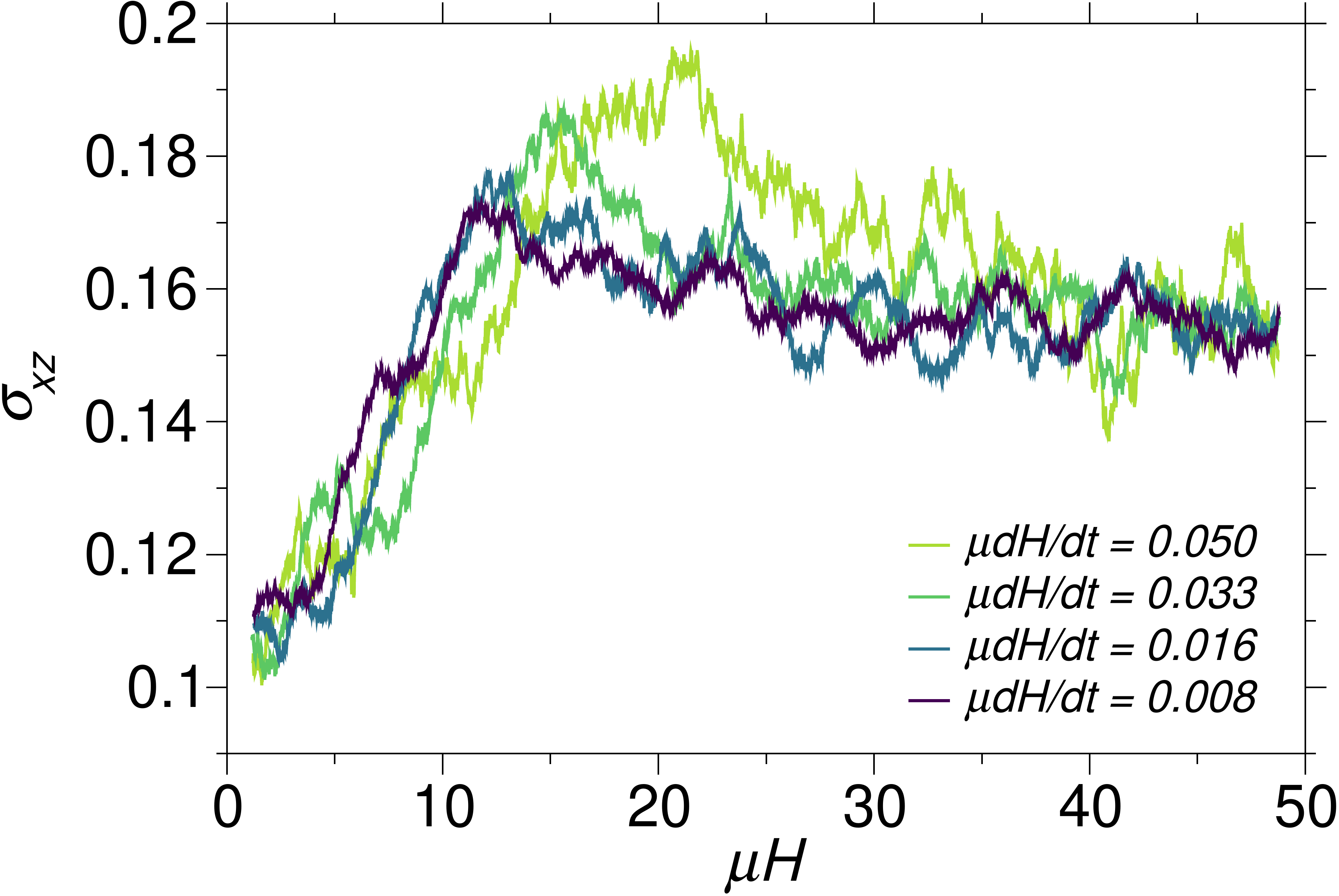}
  \caption{\label{fig:dhdt}Shear stress as a function of the magnetic field strength for different values of $\frac{ d H}{d t}$ at $T=1.5$. The results show that, in spite of quantitative differences, the qualitative dependence of the shear stress on $H$ does not depend on the $dH/dt$ value (within the considered range of values). }
\end{figure}

In this study, as mentioned in \secref{sec:simulation-setup}, the magnetic field is increased gradually from zero to $H_\mathrm{max}$ over the time interval $t_\mathrm{max}$, {i.e. with the rate of $\frac{ d H}{d t}=\frac{H_\mathrm{max}}{t_\mathrm{max}}$}. {Here we check whether the non-monotonic behavior, which is the subject of this study, is affected  by changing the values of $\frac{ d H}{d t}$.} {The magnetic field dependence of the shear stress is shown in \figref{fig:dhdt}, for a range of $\frac{ d H}{d t}$ values. The rates are chosen both larger and smaller than the rate used primarily in this work, which corresponds to $\mu\frac{ d H}{d t}=0.033$~. {S}imilar to the temperature dependence of shear stress (see \figref{fig:stress-vs-B}), {the results in \figref{fig:dhdt}} show that, although the $\sigma_{xz}$-$H$ curve changes quantitatively for different $\frac{ d H}{d t}$ values, its non-monotonic behavior is not affected at qualitative level for the examined $\frac{ d H}{d t}$ values. }
\newpage
\section{Stress auto-correlation function}
\label{sec:sacf}
\begin{figure}
 \includegraphics[width=0.45\textwidth]{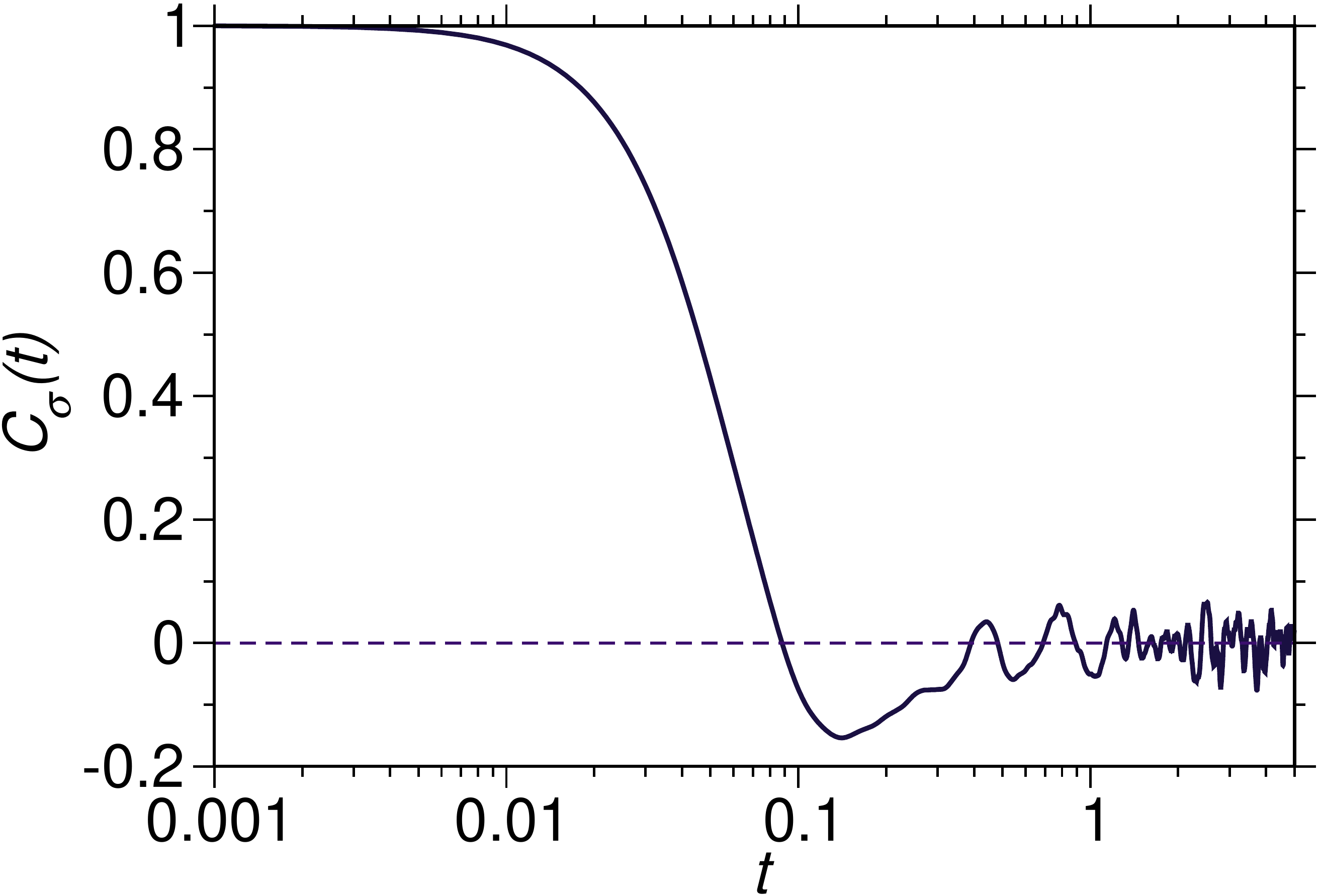}
  \caption{\label{fig:sacf} The shear stress auto-correlation function, measured at $T=1.5$ and in absence of shear, which shows that after time $t\simeq1$ the stress auto-correlation decays to zero.}
\end{figure}

Applying shear as an external perturbation leads to the response of the system in the form of an increase {of} the shear stress. In {the} liquid phase, there is a finite characteristic time, usually referred to as the relaxation time, which  the system {needs} to {ad{a}pt to} the induced stress (i.e.\ to reduce it to zero). This relaxation time is used to distinguish between high and low shear rates: if the inverse shear rate is much larger than the relaxation time, the applied shear rate is considered to be slow compared to the relaxation of the system and one would expect a linear response of the system. {Here, we calculate the relaxation time associated with stress, $\tau_\sigma$}. In order to {calculate} such a relaxation time,  we {analyze} the stress auto-correlation{, $C_\sigma(t):=\langle\sigma_{xz}(t)\sigma_{xz}(0)\rangle/\langle\sigma_{xz}^2(0)\rangle$, which is calculated} in absence of shear. {Here{,} $\langle.\rangle$ refers {to an} ensemble average.} As shown in \figref{fig:sacf}, the stress auto-correlation decays to zero and one can assign an approximate time to this decay{.} For $T=1.5$, the {obtained} relaxation time is $\tau_\sigma\simeq 1$~.

\section{Simplified Onsager analysis}\label{sec:onsager}
{In this Appendix, by using} a simplified version of the Onsager analysis presented in Refs.~\cite{onsager1949effects,lekkerkerker1984isotropic}, we show that for a system composed of hard ellipsoids of identical shapes, {an} orientational misalignment between the particles is {sufficient} to cause a demixing. More specifically, we {consider}  a system composed of two
  particle {species} $A$ and $B$, and the corresponding one{-}particle orientation distribution functions,  $f_A$ and $f_B$. {We show that} under certain constrains, the free-energy of the system is minimized if the two particle species are spatially demixed.

  For simplicity and without loss of generality, we assume that $f_A$ and $f_B$ are given by the Dirac delta distribution {located} at $\vn_A$ and $\vn_B$, {i.e. the average nematic directions}. We also assume a canonical ensemble with total {fixed } particle number  $N=N_A+N_B$,  volume $V$, and temperature $T$, where  $N_A$ and $N_B$ are the numbers of particles in species $A$ and $B$. {We are interested} in the  free energy of the system when (i) both particle species are homogeneously distributed ($F_\mathrm{hom.}$), and (ii) particles of different species are separated from each other completely ($F_\mathrm{inh.}$).

Following Ref.~\cite{lekkerkerker1984isotropic}, the reduced free energy per particle for a spatially homogeneous distribution, i.e. $\mathcal{F}_\mathrm{hom.}=F_\mathrm{hom.}/(N\kBT)$, reads 
\begin{align}\label{eq:Leker3intext}
\mathcal{F}_\mathrm{hom.}&=1+\ln(c)+(1-x)\ln(1-x)+x\ln(x) \nonumber \\ 
&\quad+(1-x)\sigma[f_A]+{x\sigma[f_B]}\nonumber \\
  &\quad +cb_\parallel(1-x)^2\rho[f_A,f_A]+cb_\parallel{x^2\rho[f_B,f_B]}\nonumber \\
  &\quad +2cb_\perp x(1-x)\rho[f_A,f_B]~,
\end{align}
where  $x$ is the fraction of $B$ particles, $c=N/V$ is the {overall} number density,  $b_\parallel=\pi/L^2D$, and $b_\perp\simeq L^2D$, with $b_\parallel$ and $b_\perp$ being the excluded volumes of two long ellipsoids of length $L$  and diameter $D$  in parallel and perpendicular configurations~\cite{onsager1949effects}. {In the above equation, the functional $\sigma[f]$  measures the entropy associated with the distribution $f$ itself, and $\rho[f,f']$ measures the entropy associated with the volume available to neighboring particles with two distributions $f$ and $f'$ (the exact expressions can be found in Onsager's work \cite{onsager1949effects}).} Assuming $L\gg D$, {that is, a needle-like shape,} $\mathcal{F}_\mathrm{hom.}$ can be approximated by
  \begin{align}\label{eq:Leker3Fhom}
    \mathcal{F}_\mathrm{hom.}&=1+\ln(c)+(1-x)\ln(1-x)+x\ln(x)\nonumber \\
                             &\quad +2cb_\perp x(1-x)|\sin(\Theta)|~,
  \end{align}
 where $\Theta$ is the angle between $\vn_A$ and $\vn_B$. {Similarly, we obtain {a}} reduced free energy for the  case where the species $A$ and $B$ are spatially separated. {In this case, the free energy per particle for each of the species is obtained by setting $x$ to zero, as each phase is purely composed of one species. This leads to }
  \begin{align}\label{eq:Leker3}
    \mathcal{F}_\mathrm{inh.}&=1+\ln(c)~,
  \end{align}
where the free-energy associated with the boundary between the two groups is neglected. 
The difference between the free energies {in the demixed state and that in the mixed state, } $\Delta \mathcal{F}=\mathcal{F}_\mathrm{inh.}-\mathcal{F}_\mathrm{hom.}$, is {given by} 
\begin{align} \label{eq:deltaF}
\Delta \mathcal{F}=(1-x)\ln(1-x)+x\ln(x)+2cb_\perp x(1-x)|\sin(\Theta)|~.
\end{align}
 The first two terms {on the right side of \equref{eq:deltaF}} are always negative and {thus} favor {a mixed system, while} the third term is always positive and {thus} favors demixing. The magnitude of the third term increases by increasing $\Theta$, which might {eventually} lead to a sign change for $\Delta\mathcal{F}$. It is straightforward to show that, depending on the values of $\Theta$ and $x$,  $\Delta\mathcal{F}$ can become positive. {In particular,} one can show that there is a critical density $c_\mathrm{cr}$ (with $c_\mathrm{cr} b_\perp=2\ln(2)$), below which always mixing is favored. 

\bibliographystyle{prsty_noetal} 
\bibliography{shorttitles,biblio}

\end{document}